\documentclass[aps,prl,twocolumn,superscriptaddress,showpacs,preprintnumbers]{revtex4-1}

\usepackage{amsmath}
\usepackage{amsfonts}
\usepackage{amssymb}
\usepackage{txfonts}
\usepackage[dvipdfmx]{graphicx}
\usepackage{color}
\usepackage[colorlinks]{hyperref}

\definecolor{CiteColor}{rgb}{0,0.5,0}
\hypersetup{citecolor=CiteColor}

\newcommand{\calE}{{\cal E}}
\newcommand{\calL}{{\cal L}}

\newcommand{\atISCO}{_{\rm isceo}}

\newcommand{\rmd}{{\rm d}}

\definecolor{red  }{rgb}{1,0,0}
\definecolor{blue }{rgb}{0,0,1}
\definecolor{green}{rgb}{0,1,0}

\newcommand{\be}{\begin{equation}}
\newcommand{\ee}{\end{equation}}
\newcommand{\bea}{\begin{eqnarray}}
\newcommand{\eea}{\end{eqnarray}}

\newcommand{\eps}{\epsilon}

\begin{document}

\title{Hamiltonian Hydrodynamics and Irrotational Binary Inspiral }

\def\addYITP{Yukawa Institute for Theoretical Physics, Kyoto university, Kyoto, 606-8502, Japan}
\def\addNagoya{Division of Particle and Astrophysical Science,
Graduate School of Science, Nagoya University,Nagoya 464-8602, Japan}
\def\addKyoto{Department of Physics, Kyoto University, Kyoto 606-8502, Japan}
\def\addSoton{School of Mathematics, University of Southampton, Southampton SO17 1BJ, United Kingdom}
\def\addShef{Consortium for Fundamental Physics, School of Mathematics and Statistics, \\ University of Sheffield, Hicks Building, Hounsfield Road, Sheffield S3 7RH, United Kingdom}
\def\addLUTh{Laboratoire Univers et Th\'eories (LUTh), Observatoire de Paris, CNRS, \\ Universit\'e Paris Diderot, 5 place Jules Janssen, 92190 Meudon, France}
\def\addRoch{Center for Computational Relativity and Gravitation, Rochester Institute \\ of Technology, Rochester, New York 14623, USA}
\def\addUCD{School of Mathematical Sciences and Complex \& Adaptive Systems Laboratory, \\ University College Dublin, Belfield, Dublin 4, Ireland}

\author{Charalampos M. Markakis}
\email{c.markakis@soton.ac.uk}
\affiliation{Theoretical Physics Institute, University of Jena, 07743 Jena, Germany}
\affiliation{Mathematical Sciences, University of Southampton, Southampton SO17 1BJ, United Kingdom}

\date{\today}


\preprint{Preprint}


\begin{abstract}
Gravitational waves from neutron-star and black-hole binaries carry valuable information on their physical properties and probe physics inaccessible to the laboratory. Although development of black-hole gravitational-wave templates in the past decade has been revolutionary, the corresponding work for double neutron-star systems has lagged. Neutron stars can be well-modelled as simple barotropic fluids during the part of binary inspiral most relevant to gravitational wave astronomy, but the crucial geometric and mathematical consequences of this simplification have remained computationally unexploited. In particular,  Carter and Lichnerowicz have described barotropic  fluid motion via   classical variational principles as conformally geodesic. Moreover, Kelvin's circulation theorem implies that initially irrotational flows  remain irrotational. Applied to numerical relativity, these concepts lead to novel  Hamiltonian or Hamilton-Jacobi schemes for evolving relativistic fluid flows. Hamiltonian  methods can conserve not only flux, but also   circulation and symplecticity, and moreover do not require addition of an artificial atmosphere typically required by standard conservative methods. These  properties can allow production of   high-precision gravitational waveforms at low computational cost. This canonical hydrodynamics approach is  applicable  to a wide class of problems  involving theoretical or computational fluid dynamics.

\end{abstract}


\maketitle

{\it Introduction}.---A wide variety of compact stellar objects where general relativistic effects are important is currently known. Black holes and neutron stars  are involved in many astrophysical phenomena, including binary mergers and gamma ray bursts, which have observable imprints in the electromagnetic and gravitational wave spectrum.
Many of these phenomena can be modelled by means of general relativistic hydrodynamics. In particular, flows describing  cosmological fluid expansion \cite{Johnston2010}, certain types of  accretion  \cite{Bondi1952,Michel1972,Font1993,Ujevic2001,Ujevic2002},  binary neutron star    
\cite{Bonazzola1997,Teukolsky1998,Shibata1998,Gourgoulhon1998,Uryu2000,Tichy2009,Taniguchi2010} or black hole-neutron star \cite{Tsokaros2007,Foucart2008,Kyutoku2009} inspiral and other phenomena, can be well-modelled as irrotational. 

With gravitational-wave astronomy about to become a reality, 
and given that  inspiral signal detection and parameter estimation typically requires  prior theoretical  knowledge of the waveforms, 
great effort has been made towards  source modelling and  
  accurate waveform template construction. Although development of black-hole gravitational wave templates in the past decade has been revolutionary, the corresponding work for  neutron-star systems has lagged in accuracy  due to the presence of matter
\cite{Read2009,Markakis2009,Markakis2010,Bernuzzi2012a,Bernuzzi2012,Read2013}.

Barotropic flows  accurately model binary neutron stars in their inspiral phase  \cite{Gourgoulhon2006,FriedmanStergioulas2013}. Synge \cite{Synge1937} and Lichnerowicz \cite{Lichnerowicz1967}  have  shown that relativistic barotropic  flows may be described via classical variational principles as  conformally geodesic. 
Carter  \cite{Carter1979}   used a non-affinely parametrized action to construct a  super-Hamiltonian and to elegantly derive covariant  4-dimensional hydrodynamic conservation laws.  

In an effort towards `clean' gravitational waveforms,
this paper outlines a canonical hydrodynamics approach that provides insight, technical simplification and gain in efficiency and accuracy to problems involving binary inspiral. To this end,   we adopt  Carter's framework but introduce  a 3-dimensional \textit{constrained} Hamiltonian  
based on an affinely parametrized action. We obtain  variational principles  in a covariant 3+1 form valid for  both Newtonian gravity and   general relativity.
Moreover, we exploit the implications of Kelvin's theorem for  relativistic irrotational hydrodynamics and  construct a  strongly hyperbolic evolution scheme with novel properties, applicable to binary neutron star inspiral and other problems. Notably, the constrained Hamiltonian approach is strictly flux-conservative and naturally 
eliminates the need for an artificial 
atmosphere, typically required by  conservative methods (see also  \cite{Millmore2010}
for a  level-set approach).
 Additionally, this approach has promising applications in theoretical \cite{Fefferman2000,Andersson2007} and computational \cite{Wilson2007,Font2008,Rezzolla2013} fluid dynamics, in a wide variety of Newtonian and relativistic contexts.  

  Below we outline our  constrained Hamiltonian formulation and its features; we relegate full details and results to a forthcoming paper. Generalization to non-irrotational or non-barotropic flows is also deferred to future work.
Spacetime indices are Greek and spatial indices Latin. We set $G = c = 1$ and   use $\nabla_\alpha$ or $\partial_\alpha$ to denote the (Eulerian) covariant  or partial derivative compatible with a curved  or flat  metric respectively, and $\partial/\partial x^\alpha$ 
to denote the (Lagrangian) partial derivative of a function $f(x,\upsilon)$ with respect to $x^\alpha$  for fixed $\upsilon^\beta$.

{\it Barotropic thermodynamics}.---Consider
a perfect fluid with  proper energy density $\eps$ and pressure $p$. 
Let us assume that the fluid is a \emph{simple barotropic fluid}, that is, all  thermodynamic quantities depend only on  rest-mass density $\rho$ and the fluid is `cold' (zero temperature) or homentropic.  For barotropic fluids, the specific enthalpy $h$ is equal to the chemical potential and satisfies the Gibbs-Duhem relation \cite{Gourgoulhon2006,FriedmanStergioulas2013,Moldenhauer2014}
\be \label{eq:relativisticenthalpyint}
h(\rho) := \frac{\eps+p}{\rho} = 1+\int_0^p \frac{dp}{\rho}=1+\eta
\ee
where $\eta$ is the kinetic specific enthalpy, satisfying   $\eta \ll 1$
in the Newtonian limit. The relation
$\rho = \rho(h)$
 is  the equation of state (EOS)  of the fluid.

{\it Euler-Lagrange hydrodynamics}.----To set the stage, we review  Euler-Lagrange dynamics in covariant language as applied to  fluid theory in Newtonian gravity or 3+1 general relativity    (the relativistic four-dimensional formulation is outlined in \cite{Carter1979} and its  generalization to magnetohydrodynamics is given in \cite{Markakis2011a}). The results derived in this paper will apply to any motion
in which the flow lines obey a Lagrangian variation
principle. That is, for any particular flow configuration,
there exists a Lagrangian function $L(t,x,\upsilon)$ of the spacetime coordinates $x^\alpha=\{t,x^a\}$ and \textit{canonical 3-velocity} ${\upsilon^a} = d{x^a}/dt$ of a fluid element measured in local coordinates. 
Consider a fluid element of unit mass moving along a streamline under the influence of pressure and gravitational forces. We assert  that, in both nonrelativistic and relativistic contexts, and for both self-gravitating or test fluids, the motion of a fluid element can be obtained from   an action of the form
\be \label{FluidActionNewt}
S = \int_{{t_1}}^{{t_2}} {L(t,x,\upsilon)dt}   
\ee
Minimizing the action yields the 
Euler-Lagrange equation of motion:
\be  \label{ELeqsNewt}
{{d{p_a}} \over {dt}} - {{\partial L} \over {\partial {x^a}}} = ({\partial _t} + {\pounds_\upsilon}){p_a} - {\nabla _a}L = 0
\ee
where ${p_a} = \partial L(t,x,\upsilon)/\partial {\upsilon^a} $
is the \textit{canonical momentum}   of the fluid element conjugate to $x^a$, $\pounds_\upsilon$ is the Lie derivative along $\upsilon^a$ and 
${\nabla _a}L := {\partial}L(t,x,\upsilon)/\partial x^a +p_b \partial{\upsilon ^b}/\partial x^a$. 
As emphasized by Carter  \cite{Carter1979}, the second, covariant version of Eq.~\eqref{ELeqsNewt} is the form appropriate in a fluid-theory context.

For a barotropic fluid,  Eq.~\eqref{eq:relativisticenthalpyint} implies that the pressure force arises from a potential.  Then,  the nonrelativistic Lagrangian
\be \label{nonrelativisticLagrangian}
L(t,x,\upsilon)={\textstyle{1 \over 2}}{\gamma _{ab}(x)}{\upsilon^a}{\upsilon^b} - \Phi(t,x)- \eta(t,x)   
\ee
(where $\gamma _{ab}$ is the Euclidian 3-metric and $\Phi$ is the Newtonian potential), implies
\be \label{canonicalMomentumNonrelativistic}
{p_a} =\frac{\partial L}{\partial \upsilon^a}= \gamma _{ab}{\upsilon^b}={\upsilon_a}
\ee
and, when substituted into 
Eq.~\eqref{ELeqsNewt}, yields the nonrelativistic Euler equation in Lagrangian form:
\be  \label{EulerEqNewtLagr}
({\partial _t} + {\pounds_\upsilon}){\upsilon_a} = {\nabla _a}({\textstyle{1 \over 2}}{\upsilon^2} - \Phi- \eta   ) 
\ee 
where $\upsilon^2=\upsilon_b \upsilon^b$. Barotropic fluid motion may thus be described as  motion in an effective potential $\Phi+\eta$.
 In the pressureless (`dust') limit, $\eta$ vanishes and the motion  reduces to that of a particle in a Newtonian potential $\Phi$.
 
An analogous result holds in general relativity:
barotropic fluid streamlines are  \textit{geodesics} of a   Riemannian manifold  with  metric $h^2 g_{\alpha \beta}$ \cite{Lichnerowicz1967}. These geodesics  minimize the arc length $S=  - \int_{{\tau _1}}^{{\tau _2}} {h(x)\sqrt { - {g_{\alpha \beta }}(x){u^\alpha }{u^\beta }} d\tau } $, where $g_{\alpha \beta}$ is the spacetime metric, $u^\alpha=dx^\alpha/d \tau=u^t dx^\alpha /dt$ is the fluid 4-velocity
and $\tau$ is the proper time of an observer comoving with the fluid \cite{Moldenhauer2014}.  With the standard $3+1$ decomposition, the spacetime
$\mathcal{M}\ = \mathbb{R} \times \Sigma$  is foliated by a family of spacelike surfaces $\Sigma_t$ and, in a chart $\{t,x^i\}$, its metric takes the form
$d\tau^2=-g_{\mu \nu}dx^\mu dx^\nu=\alpha^2 dt^2 -  \gamma_{ij}(dx^i+\beta^i dt)(dx^j+\beta^j dt),$
where $\alpha$ is the lapse, $\beta^a$ is the shift vector and  $\gamma_{ab}$ is the spatial metric. 
Substituting the 3+1 metric into the action $S$ and using the  coordinate time $t$ as  affine parameter leads to an action of the form~\eqref{FluidActionNewt}
with the relativistic Lagrangian given by
\be \label{relativisticLagrangian}
L(t,x,\upsilon)=-\alpha(t,x) h(t,x) \sqrt{1-{\gamma _{ab}(t,x)}{\nu^a}{\nu^b} }  =-h/u^t, 
\ee
where $\nu^a=\alpha^{-1}(\upsilon^a+\beta^a)$ is the fluid 3-velocity   measured by \textit{normal observers}. The canonical 3-momentum is given by
\be \label{relativisticMomentum}
p_a=\frac{\partial L}{\partial \upsilon^a}=h \frac{\nu_a}{ \sqrt{1-{\nu^2} }}=hu_a,
\ee
where $\nu_a=\gamma_{ab} \nu^b$ and $\nu^2=\nu_b\nu^b$. 
Substituting Eqs.~\eqref{relativisticLagrangian} and \eqref{relativisticMomentum} into Eq.~\eqref{ELeqsNewt} yields the relativistic Euler equation in Lagrangian 3+1 form:
\be  \label{EulerEqRelativisticLagr}
({\partial _t} + {\pounds_\upsilon}){(hu_a)} = {-\nabla _a}(h/u^t). 
\ee 
This equation could have  been obtained directly by  3+1 decomposing the relativistic Euler equation in four-dimensional Lagrangian form,  $\pounds_u (h u_\alpha)=-\nabla_\alpha h$
  \cite{Carter1979,Gourgoulhon2006,FriedmanStergioulas2013}, but its derivation from a  variational principle is essential for what follows. In the pressureless  limit, $h=1$, the motion  reduces to a geodesic of   $\mathcal{M}$. 
 Eqs.~\eqref{EulerEqNewtLagr} and \eqref{EulerEqRelativisticLagr}
are suitable for numerical evolution in  Lagrangian coordinates, such as smoothed-particle hydrodynamics. The  canonical approach outlined below is  suited to methods based on either Eulerian or Lagrangian coordinates.

{\it Hamiltonian hydrodynamics}.---Using
the covariant Euler-Lagrange equation \eqref{ELeqsNewt} and the  Cartan identity ${\pounds_\upsilon}{p_a} = {\upsilon ^b}({\nabla _b}{p_a} - {\nabla _a}{p_b}) + {\nabla _a}({\upsilon^b}{p_b})$, one obtains the covariant Hamilton equation
\be \label{canonicalEqs}
\frac{{d{p_a}}}{{dt}} + \frac{{\partial H}}{{\partial {x^a}}} = {\partial _t}{p_a} + {\upsilon^b}({\nabla _b}{p_a} - {\nabla _a}{p_b}) + {\nabla _a}H = 0
\ee
where
\be \label{HamiltonianLegendre}
H(t,x,p)=\upsilon^a p_a-L(t,x,\upsilon)
\ee
is the Hamiltonian of a fluid element.
Note that, like Eq.~\eqref{ELeqsNewt}, Eq.~\eqref{canonicalEqs} is  valid in  Newtonian \textit{and} relativistic contexts.

For nonrelativistic barotropic flows, Eqs.~\eqref{nonrelativisticLagrangian}, \eqref{canonicalMomentumNonrelativistic} and \eqref{HamiltonianLegendre} yield
the  Hamiltonian 
\be \label{HamiltonianNewtonian}
H(t,x,p)={\textstyle{1 \over 2}}{\gamma^{ab}(x)}{p_a}{p_b} + \Phi(t,x)+ \eta(t,x)
\ee
and Eq.~\eqref{canonicalEqs} yields the nonrelativistic
Euler equation in canonical form, also known as the Crocco equation:
\be \label{canonicalEqsNonrelativistic}
 {\partial _t}{\upsilon _a} + {\upsilon^b}({\nabla _b}{\upsilon_a} - {\nabla _a}{\upsilon_b}) + {\nabla _a}({\textstyle{1 \over 2}} {\upsilon^2} + \Phi + \eta ) = 0.
\ee

Multiplying this equation 
by the density $\rho$ and using the Gibbs-Duhem relation~\eqref{eq:relativisticenthalpyint} and the nonrelativistic continuity equation
 \be
 \partial_t \rho +\nabla_a (\rho \upsilon^a)=0
 \ee
leads to a flux-conservative form of the Euler equation:
\be
{\partial _t}{(\rho \upsilon _a)} +  {\nabla _b}{T_{a}}^{b} = {-\rho\nabla _b}\Phi.
\ee
where ${T_{a}}^{b}=\rho\upsilon_a \upsilon^b + p {\gamma_{a}}^{b}$ is the fluid stress tensor.

For relativistic barotropic flows, Eqs.  \eqref{relativisticLagrangian}, \eqref{relativisticMomentum} and \eqref{HamiltonianLegendre} yield
the  constrained Hamiltonian 
\be \label{HamiltonianRelativistic}
H(t,x,p)\!=\! - {p_a}{\beta ^a (t,x)} + \alpha(t,x) \sqrt {{h(t,x)^2} \!+ {\!\gamma ^{ab}(t,x)}{p_a}{p_b}}\!=\!-h  u_t
\ee
and Eq.~\eqref{canonicalEqs} yields the relativistic
Euler equation in 3+1 canonical form
\be \label{canonicalEqsRelativistic}
 {\partial _t}{(hu_a)} + {\upsilon^b}[{\nabla _b}{(h u_a)} - {\nabla _a}{(h u_b})] - {\nabla _a}(hu_t) = 0.
\ee
This equation could have been obtained by  3+1 decomposing the Euler equation in Carter-Lichnerowicz form, written as ${u^\beta}[{\nabla _\beta}{(h u_\alpha)} - {\nabla _\alpha}{(h u_\beta})]=0$
 in four dimensions \cite{Synge1937,Lichnerowicz1967,Carter1979,Gourgoulhon2006}.
The  Hamiltonian~\eqref{HamiltonianRelativistic} amounts to  the energy of a fluid element  measured in local coordinates and could have alternatively been obtained by solving the constraint $g^{\alpha \beta}u_\alpha u_\beta=-1$ for $u_t$. In the pressureless  limit, $h=1$, Eq.~\eqref{HamiltonianRelativistic} reduces to the constrained Hamiltonian of a particle of unit mass moving on a spacetime geodesic \cite{Barausse2009} and Eq.~\eqref{canonicalEqsRelativistic} describes a congruence of such geodesics.  

Multiplying Eq.~\eqref{canonicalEqsRelativistic}   
by the  density $\rho$ and using the Gibbs-Duhem relation~\eqref{eq:relativisticenthalpyint} and the relativistic continuity equation
\be \label{RelativisticContinuity}
\nabla_\alpha(\rho u^\alpha)=\frac{1}{\sqrt{-g}}\partial_\alpha(\sqrt{-g} \,\rho u^\alpha)=0
\ee
(where $g=\det(g_{\mu \nu})$) implies that the divergence of the fluid energy-momentum tensor  $\, {T_{\alpha}}^{\beta}=\rho h u_\alpha u^\beta + p\, {g_{\alpha}}^{\beta}$ vanishes:
\be \label{ValenciaT}
\nabla_\beta {T_{\alpha}}^{\beta} =\frac{1}{\sqrt{-g}}\partial_\beta(\sqrt{-g} \, {T_{\alpha}}^{\beta})-\Gamma^{\gamma}_{\alpha \beta} {T_{\gamma}}^{\beta}=0. 
\ee
The above flux-conservative form of the Euler equation is typically used in numerical simulation via shock-capturing methods. However,  the canonical form \eqref{canonicalEqs} carries unique advantages, especially in the irrotational case discussed below. 

\textit{Conservation of circulation}.---The canonical vorticity 2-form, $\omega_{ab}:=\nabla_a p_b-\nabla_a p_b$, satisfies an  evolution equation,
$({\partial _t} + {\pounds_\upsilon}){\omega_{ab}} = 0$,
obtained from the exterior derivative of Eq.~\eqref{ELeqsNewt}. The integral form of this equation constitutes Kelvin's circulation theorem: the circulation along a fluid ring 
$\mathcal{C}_t=\partial \mathcal{S}_t$
dragged along by the flow is conserved:
 \bea \label{KelvinTheorem}
\frac{d}{dt}\! \oint_{{\mathcal{C}_t}} \!{p_a dx^a}\!=\!\frac{d}{dt} \!\int_{{\mathcal{S}_t}} \!{\omega_{ab} \,
d \Sigma^{ab}}
\!=\! \int_{{\mathcal{S}_0}}\! {({\partial _t} + {\pounds_\upsilon}){\omega_{ab}} \,d \Sigma^{ab}}
\!=\!0 \quad \; 
\eea
where the first equality follows from the  Stokes theorem and represents the flux of vorticity through the surface $\mathcal{S}_t=\Psi_t \mathcal{S}_0$,
where $\Psi_t$ is the family of diffeomorphisms generated by fluid velocity  $\upsilon^a$.

From a general variation of the action~\eqref{FluidActionNewt}, it is possible to show \cite{Arnold1989, Boccaletti2003} that the integral
\be \label{PoincareCartan}
I= \oint_{{\mathcal{C}}} {(p_a dx^a-Hdt)},
\ee
calculated along an arbitrary closed contour $\mathcal{C}=\partial \mathcal{S}$ lying on the hypersurface $\mathcal{S}$ (to which the fluid motion is restricted) of the extended phase space $(x^a,p_a,t)$, is invariant under an arbitrary displacement or deformation of the contour along any tube of   fluid streamlines (or particle trajectories in the pressureless limit). A dynamical system admits an invariant $I$,  known as the \textit{Poincar\'e-Cartan integral invariant},  iff it  is Hamiltonian. If we consider curves $\mathcal{C}_t$ lying in planes of constant $t$ in phase space, then $dt=0$ along such curves and $I$ reduces to the conserved circulation integral in \eqref{KelvinTheorem}. In four dimensional general relativity, one typically evaluates the  integral~\eqref{PoincareCartan} along a  fluid ring $\mathcal{C}_\tau$ of constant proper time $\tau$ and writes Kelvin's theorem in the form  ${\textstyle{d \over d \tau}} \oint_{{\mathcal{C}_\tau}} {p_\alpha dx^\alpha}=0$; this conservation law can be derived directly from the relativistic Euler equation \cite{FriedmanStergioulas2013}. We stress, nevertheless, the   fact that  Eqs.~\eqref{KelvinTheorem}, \eqref{PoincareCartan} are  valid as written in both Newtonian gravity and 3+1 general relativity.

The most interesting feature of Kelvin's  theorem  is that, since its derivation did not depend on the metric, it is
exact in time-dependent spacetimes, with gravitational  waves
carrying energy and angular momentum away from a system. In
particular, \textit{oscillating stars and radiating binaries, if modeled as
 barotropic fluids with no viscosity or  dissipation other than gravitational radiation, \textit{exactly}
conserve circulation} \cite{FriedmanStergioulas2013}. 
An important  corollary of Kelvin's theorem is that, if circulation is zero initially, it must remain zero subsequently. That is, \textit{flows initially irrotational remain irrotational}. Apart from an application to incompressible Newtonian binaries  \cite{Eriguchi1990}, this concept has remained unexploited in  simulations of binary inspiral, despite the fact that numerical relativity simulations typically begin with irrotational neutron-star initial data \cite{Bonazzola1997,Teukolsky1998,Shibata1998,Gourgoulhon1998,Uryu2000,Gourgoulhon2006,Tichy2009,Taniguchi2010,Tsokaros2007,Foucart2008,Kyutoku2009}, as spin is usually negligible in this regime.  The implications of this corollary for  relativistic fluid dynamics    are explored below.

\textit{Irrotational Hamilton-Jacobi hydrodynamics}.---A flow  is called irrotational if the vorticity 2-form $\omega_{ab}$ vanishes 
\be \label{IrrotationalConstraint}
\nabla_a p_b-\nabla_a p_b=0
\ee
or, by virtue of the Poincar\'e lemma (for simply connected manifolds), if the canonical momentum is the gradient of a velocity potential:
\be \label{momentumIrrotational}
p_a=\nabla_a S(t,x)
\ee
  For irrotational flows, the Hamilton equation \eqref{canonicalEqs} simplifies to the strictly flux-conservative canonical equation
\be \label{HJgradient}
{\partial _t}{p_a} + {\nabla _a}H(t,x,p) = 0
\ee
Substituting Eq.~\eqref{momentumIrrotational} into \eqref{HJgradient} gives the first integral
\be \label{HJ}
{\partial _t}{S(t,x)} + H(t,x,\nabla S) = 0
\ee
which has the form of a Hamilton-Jacobi equation. (The integration constant $c(t)$ is eliminated by  adding $\int^t c(t')dt'$ to $S$    without altering $p_a$.)
$H$ and $p_a$ are given by Eqs.~\eqref{HamiltonianNewtonian}, \eqref{canonicalMomentumNonrelativistic} for Newtonian gravity or  Eqs.~\eqref{HamiltonianRelativistic}, \eqref{relativisticMomentum} for 3+1 general relativity.

 The above corollary to Kelvin's theorem suggests that irrotational initial data may be evolved by solving either the Hamilton-Jacobi equation \eqref{HJ} or its gradient, the Hamilton equation~\eqref{HJgradient}.
This is \textit{equivalent} to solving the  Euler equation -- there is no approximation involved -- as long as the initial data is irrotational. (In fact,  even for a non-barotropic EOS, the Euler equation may be used to show that  initially irrotational flows are also initially barotropic, i.e. homentropic or zero temperature, and remain so subsequently). 

 For the Hamiltonian functions given above, Eq.~\eqref{HJgradient} can be considered a generalization of the Burgers
equation. In the absence of pressure and gravitational forces, by virtue of Eqs.~\eqref{HamiltonianNewtonian} and \eqref{canonicalMomentumNonrelativistic},   Eq.~\eqref{HJgradient}
reduces to the nonrelativistic Burgers equation, 
$ 
{\partial _t}{\upsilon_a} + {\partial _a}({\textstyle{1 \over 2}} {\upsilon^2}  ) = 0
$.
In Minkowski space, by virtue of Eqs.~\eqref{HamiltonianRelativistic} and \eqref{relativisticMomentum}, Eq.~\eqref{HJgradient}
similarly reduces to a special-relativistic  Burgers equation, 
$ 
\partial_t (\upsilon_a/ \sqrt{1-\upsilon^2}) + \partial _a(\sqrt{1+\upsilon^2}) = 0 $,
which  reduces to the nonrelativistic equation for  $\upsilon \ll 1$.
LeFloch et al.  \cite{LeFloch2012a,LeFloch2012} provide a non-covariant derivation of  this equation for Minkowski and Schwarzschild spacetimes, based on algebraic manipulation of the Euler and continuity equations on  particular charts rather than covariant variational principles; numerical evolutions of these equations in 1+1 dimensions were successful, even in the presence of shocks. However, the  fact that such equations amount to  Hamilton or Hamilton-Jacobi equations,  that can be obtained from constrained particle-like variational principles and written  in  covariant 3+1 form   for \textit{any} spacetime,
 remains  unnoticed.
The covariant approach outlined above motivates  the use of  Eqs.~\eqref{relativisticMomentum}, \eqref{HamiltonianRelativistic} and \eqref{HJgradient} or \eqref{HJ} for  irrotational hydrodynamics in a  variety of physical contexts. 

Several  methods (cf.~\cite{Jin1998,Jiang2000,Bryson2003,Qiu2005,Carlini2006,Zhang2006} and references therein) exist for solving  Hamilton-Jacobi equations  numerically.  A well-known mathematical problem encountered with such  equations is non-uniqueness of solutions, but    unique `viscosity solutions' may be obtained in the limit of small viscosity     \cite{Crandall1992b}. Eq.~\eqref{HJ} provides the possibility of applying such well-established methods in the context of Newtonian or relativistic   fluid dynamics. Although this equation has the advantage of being scalar, there are  certain advantages to   using its  flux-conservative  canonical form  \eqref{HJgradient} for computational purposes. In the latter approach, one may make use of existing flux-conservative scheme, abundantly implemented in numerical relativity, without  artificial viscosity, but must check
that the constraint \eqref{IrrotationalConstraint} is satisfied;  such violations also appear in  the standard  approach and may be eliminated via relaxation techniques \cite{Jin1998}. The canonical equation   \eqref{HJgradient} is coupled, via a barotropic equation of state $\rho=\rho(h)$, to the continuity equation. In general relativity, the latter is given by Eq.~\eqref{RelativisticContinuity}   and can be decomposed as
\be \label{RelativisticContinuity3p1}
\partial_t \rho_{\star} +\partial_a (\rho_{\star} \upsilon^a)=0
\ee
where $\rho_{\star}:=\sqrt{-g} \, \rho u^t=\alpha\sqrt{\gamma} \, \rho u^t$  and $\gamma=\det(\gamma_{ij})$. 
Then, the system of Eqs.~\eqref{HJgradient} and \eqref{RelativisticContinuity3p1}
can be written as
\be \label{FluxConservation}
{\partial _t}{\bf{U}} + {\partial _k}{{\bf{F}}^k} = 0
\ee
where the components of the   conservative variable vector $\bf{U}$  and flux vectors  ${{\bf{F}}^k}$   are given by
\be
{\bf{U}} = \left( {\begin{array}{*{20}{c}}
{{\rho_\star}}\\
{{p _i}}
\end{array}} \right),\quad {{\bf{F}}^k} = \left( {\begin{array}{*{20}{c}}
{{\rho_\star}{\upsilon^k}}\\
{\delta _i^kH}
\end{array}} \right),\quad k = 1,2,3
\ee
and $p_i,H$   are given by Eqs.~\eqref{relativisticMomentum},  \eqref{HamiltonianRelativistic}. In the Newtonian limit, one sets $\rho_{\star}:=\sqrt{\gamma} \, \rho$ and uses Eqs.~\eqref{canonicalMomentumNonrelativistic},  \eqref{HamiltonianNewtonian} instead.

Eq.~\eqref{FluxConservation}  can be evolved together with the spacetime metric \cite{Alcubierre2008,Bona-Casas2009,Baumgarte2010a} and
 is our main result.
  Notably, this evolution system is  source-free, and thus \textit{strictly} flux-conservative, with no further assumptions such as Killing symmetries.
Moreover, for finite sound speed $c_{\rm{s}}\!=\! \! \sqrt{dp/d \epsilon} \!=\! \! \sqrt{d \ln h/ d \ln \rho}$, the  system is \textit{strongly hyperbolic} and thus has a well-posed initial value problem: a lengthy but straightforward characteristic analysis   
 shows that the   system 
possesses a complete basis of four eigenvectors, with eigenvalues  
$\lambda^k_{1,2}=0$ (double) and $\lambda^k_{3,4}=\alpha ({1 - {\nu ^2}c_{\rm{s}}^2})^{-1} \{ \nu ^k(1 - c_{\rm{s}}^2) 
\pm c_{\rm{s}}^{} {(1 - {\nu ^2})^{1/2}[(1 - {\nu ^2}c_{\rm{s}}^2){\gamma ^{kk}} - (1 - c_{\rm{s}}^2){{({\nu ^k})}^2}]^{1/2}} \}  - {\beta ^k} $.
The latter pair of `acoustic' eigenvalues is identical to those of the Valencia formulation, while the former pair is different \cite{Banyuls1997}. 
 When numerically evolving Eq.~\eqref{FluxConservation}, one needs to construct the fluxes   ${{\bf{F}}^k}$   given the conserved variables ${\bf{U}}$  
at each time step. To do so, one  needs to
recover the primitive variables $\{h,u_i\}$ given ${\bf{U}}$,  by first  solving for $h$ the algebraic equation
\be
\rho (h) = \frac{{\rho_{\star}h}}{{\sqrt \gamma  \sqrt {{\gamma ^{ij}}{p_i}{p_j} + {h^2}} }}
\ee
for fixed $\rho_{\star}$,  $p_i$ and $\gamma_{ij}$. This equation is obtained by substituting the relation  ${u^t} = {\alpha ^{ - 1}}\sqrt {{\gamma ^{ij}}{u_i}{u_j} + 1}$ into the definition of $\rho_\star$ and using  Eq.~\eqref{relativisticMomentum}.   A novel feature of  Eq.~\eqref{FluxConservation}   is that the recovery of    $u_i$ is performed by dividing $p_i$  by the specific enthalpy $h$
which becomes unity on the  surface, rather than dividing    $\rho_\star u_i$ by the density  $\rho$ which vanishes there.
Thus, unlike the standard approach, \textit{no artificial atmosphere} \textit{is required for recovery of primitive from conservative variables}. 

\textit{Conclusions}.---Although the Carter-Licherowicz approach \cite{Lichnerowicz1967,Carter1979} has been used to obtain first integrals for constructing initial data for compact binaries in the presence of Killing symmetries    
\cite{Bonazzola1997,Teukolsky1998,Shibata1998,Gourgoulhon1998,Uryu2000,Gourgoulhon2006,Tichy2009,Taniguchi2010,Tsokaros2007,Foucart2008,Kyutoku2009}, it has never been adopted to   fluid flow evolution. 
Moreover, since irrotationality is independent of helical symmetry, this simplification applies not only to circular but also inspiralling or eccentric nonspinning binaries, but has yet to be exploited in hydrodynamic simulations. This paper provides the steps  towards  these goals. Numerical tests of the irrotational hydrodynamics system~\eqref{FluxConservation}  have been performed successfully;  details and results from simulation of binary neutron star inspiral will be provided in a future paper.

Avoiding an artificial atmosphere does not only increase  accuracy (as systematic errors related to the atmosphere are eliminated) but also   increases efficiency (as  numerical operations  for hydrodynamics outside the star are avoided).
As mentioned earlier, unlike the energy-momentum conservation laws \eqref{ValenciaT}, the  irrotational conservation laws~\eqref{FluxConservation}  are source-free and represent \textit{strict} conservation. This feature simplifies implementation and increases precision as it avoids numerical differentiation of the metric. A caveat is that the Hamiltonian is nondifferentiable at the star surface, so  care must be taken in performing numerical differentiation at that location to retain accuracy, as detailed elsewhere.
Additional accuracy can be gained by using \textit{symplectic integration} schemes for time evolution,  that preserve  Hamiltonian structure and circulation.

Finally, although the  above approach     focused on irrotational flows,  it is feasible to accommodate non-irrotational or even non-barotropic flows in the formulation while retaining most of its merits.   Such developments are expected to be of interest in theoretical and computational fluid dynamics, in  Newtonian and relativistic contexts, and motivate future work.


\acknowledgments
The author  thanks Nils Andersson, Sebastiano Bernuzzi, Bernd Br\"ugmann, Brandon Carter, Kyriaki Dionysopoulou, John L. Friedman, Eric Gourgoulhon, Carsten Gundlach, Ian Hawke, David Hilditch, Koutarou Kyutoku, Niclas Moldenhauer, Frans Pretorius, Gerhard Sch\"afer, Masaru Shibata,
Kostas Skenderis, Branson Stephens, Nikolaos Stergioulas, Keisuke Taniguchi and K\=oji Ury\=u for their valuable feedback.
The author gratefully acknowledges  support by the DFG SFB/Transregio 7 
``Gravitational Wave Astronomy'',  the  STFC grant PP/E001025/1 as well as the hospitality of LUTh - Observatoire de Paris where part of this work was completed.

\bibliography{library}


\end{document}